# Coverage Enhancement Strategy in WMSNs Based on a Novel Swarm Intelligence Algorithm: Army Ant Search Optimizer

Yin-Di Yao, Qin Wen, *Student Member, IEEE*, Yan-Peng Cui, *Student Member, IEEE*, Feng Zhao, Bo-Zhan Zhao, and Yao-Ping Zeng

*Abstract*—As one of the most crucial scenarios of the Internet of Things (IoT), wireless multimedia sensor networks (WMSNs) pay more attention to the information-intensive data (e.g., audio, video, image) for remote environments. The area coverage reflects the perception of WMSNs to the surrounding environment, where a good coverage effect can ensure effective data collection. Given the harsh and complex physical environment of WMSNs, which easily form the sensing overlapping regions and coverage holes by random deployment. The intention of our research is to deal with the optimization problem of maximizing the coverage rate in WMSNs. By proving the NP-hard of the coverage enhancement of WMSNs, inspired by the predation behavior of army ants, this article proposes a novel swarm intelligence (SI) technology army ant search optimizer (AASO) to solve the above problem, which is implemented by five operators: army ant and prey initialization, recruited by prey, attack prey, update prey, and build ant bridge. The simulation results demonstrate that the optimizer shows good performance in terms of exploration and exploitation on benchmark suites when compared to other representative SI algorithms. More importantly, coverage enhancement AASO-based in WMSNs has better merits in terms of coverage effect when compared to existing approaches.

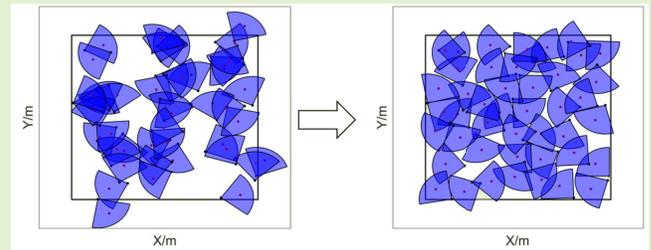

## I. Introduction and Related Works

INTERNET of Things (IoT)-based monitoring of various environmental parameters and resources is growing in domains and dimensions day by day [1]. The paradigm of IoT involves a large number of smart devices, which are capable of accessing data about the environment in a remote field cooperatively and intelligently without any human interaction [2]. Wireless sensor networks (WSNs) have been considered the key enablers of IoT with the development of intelligent sensor technology, distributed information processing technology, and computational vision [3]. WSNs have gradually attracted significant attention because of their wide application such as environmental monitoring and precision agriculture [4], [5].

As the monitoring environment becomes increasingly complex and changeable, the simple data (e.g., temperature, humidity, and light intensity) obtained by traditional WSNs no longer satisfy the monitoring needs [6]. With the development of image/video sensors, ultrasonic sensors, and infrared sensors, directional sensor networks (DSNs) have attracted great attention from academia and industry. To characterize the directional sensing feature of nodes, the sensing range of a directional sensor is limited by visual angle. As an important application of DSNs, the wireless multimedia sensor network (WMSN) is a technology where the multimedia sensor nodes (e.g., camera, microphone) with computing, storage, and communication capabilities can sense, fuse, and process multiple

Manuscript received 28 June 2022; revised 24 August 2022; accepted 24 August 2022. Date of publication 15 September 2022; date of current version 31 October 2022. This work was supported in part by the National Nature Science Foundation of China under Grant U1965102, in part by the Science and Technology Innovation Team for Talent Promotion Plan of Shaanxi Province under Grant 2019TD-028, in part by the Agricultural Project of Science and Technology Department of Shaanxi Province under Grant 2021NY-180, in part by the Communication Soft Science Project of the Ministry of Industry and Information Technology under Grant 2021-R-47, and in part by the Key Research and Development Project of Shaanxi Province under Grant 2020NY-161. The associate editor coordinating the review of this article and approving it for publication was Prof. Jari Nurmi. *(Corresponding author: Qin Wen.)*

Yin-Di Yao, Qin Wen, Bo-Zhan Zhao, and Yao-Ping Zeng are with the School of Communication and Information Engineering and the School of Artificial Intelligence, Xi'an University of Posts and Telecommunications, Xi'an 710121, China (e-mail: yaoyindi@xupt.edu.cn; wq199802@163.com; zhaobozhan1999@stu.xupt.edu.cn; cengyaoping@xupt.edu.cn).

Yan-Peng Cui is with the Key Laboratory of Universal Communications, Ministry of Education, Beijing University of Posts and Telecommunications, Beijing 100876, China (e-mail: cuiyanpeng94@bupt.edu.cn).

Feng Zhao is with the School of Cyberspace Security, Xi'an University of Posts and Telecommunications, Xi'an 710121, China (e-mail: zhaofeng2@xupt.deu.cn).

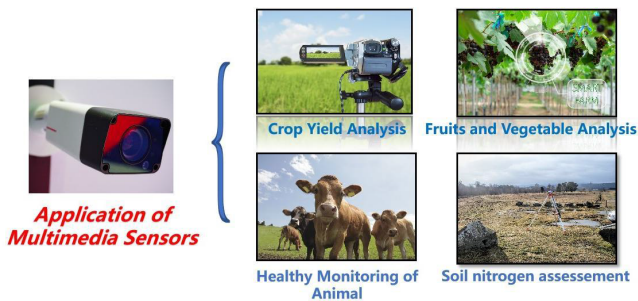

Fig. 1. Application of multimedia sensors.

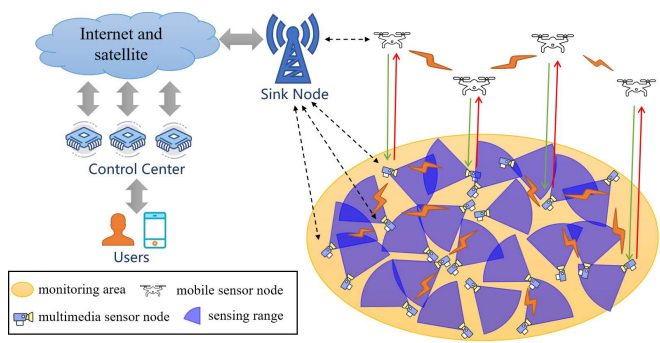

Fig. 2. Coverage model.

media information (e.g., audio, video, image, and sonar) in the surrounding environment [7]. In Fig. 1, we can see that multimedia sensors with advanced artificial intelligence, computer vision, and big data analysis are applied in many fields [8], [9]. For multimedia sensor nodes operating in a monitoring environment, as shown in Fig. 2, the collected multiple media information can be transmitted to the sink node based on the preset routing protocol [10] or mobile devices (e.g., unmanned aerial vehicles, mobile robots) [11]. The multimedia sensor nodes can harvest energy from radio-frequency signals transmitted by dedicated energy sources and realize local processing and/or task unloading of multimedia information based on an advanced resource allocation algorithm [12].

One important issue of WMSNs is coverage control, which reflects the perception of the surrounding environment and the quality of service of WMSNs, where a good coverage effect can ensure effective data collection [13]. Given the harsh and complex monitoring environment, a stochastic deployment is a viable approach to the deployment. WMSNs are usually randomly deployed from the low-flying plane pallet, artillery shell, and missiles to obtain situation awareness in the monitoring area [14]. However, it is difficult to guarantee the coverage area in this fashion. Additionally, considering the limitation of the energy supplement and recovery of multimedia sensor nodes, the position of multimedia sensor nodes becomes invariable once deployed, but its sensing direction is adjustable [15]. Therefore, the intention of our research is to satisfy the coverage requirements by finding the optimal sensing direction of a set of multimedia sensor nodes, and such a procedure is called the coverage enhancement problem of WMSNs (CEPW).

Considering that WMSNs is an important branch of DSNs, the coverage enhancement of DSNs will be described next. The existing research on coverage enhancement mainly includes the following two: coverage enhancement strategy based on geometry and coverage enhancement strategy based on swarm intelligence (SI).

As the most classical coverage enhancement algorithm, the essence of the virtual force algorithm (VFA) is to separate overlapping sensors and fill sensing blind areas. A large number of algorithms have been derived from VFA. A potential field-based coverage-enhancing algorithm (PFCEA) is proposed in [16], and CEPW is transformed into a uniform distribution problem of centroids by defining the repulsion force between directional sensor nodes, which enhances the coverage effect effectively. In order to shut off the redundant sensors and enhance the coverage effect of WMSNs, the work of [17] proposed a virtual centripetal force-based coverage-enhancing algorithm for WMSNs by introducing the approach of visual potential and the theory of grid. Ma *et al.* [18] proposed a 3-D directional sensor coverage-control model with tunable orientations and developed the virtual repulsion between neighbor nodes to adjust the horizontal and vertical offsets angles, which realized the coverage enhancement of the projection area of the sensor. Li *et al.* [19] proposed a coverage enhancement method based on the Delaunay to strengthen the coverage of weak points and improve the coverage of the hot region, in which sensors rotated their directions toward the nearby weak spots. Utilizing the Voronoi diagram in [20], the authors enhanced the coverage effect in WMSNs. The work of [21] proposed the least overlapped-area first with a rotatable algorithm and the update priority with a rotatable algorithm to repair the coverage holes of WMSNs, which improved the coverage rate on different parameters of the sensor.

SI has been developed rapidly due to its simplicity, strong adaptability, and high stability [22], [23]. The application of SI in coverage enhancement has become the focus of scholars [24], [25], [26]. The coverage problem of DSNs is solved by researchers with particle swarm optimization (PSO), which is one of the most classical SI algorithms. To reduce the redundancy ratio and improve the coverage rate caused by the random deployment, an improved adaptive PSO (IAPSO) was proposed in [27] by establishing the optimization model of the coverage problem of DSNs. In literature [28], a virtual angle boundary-aware PSO is designed to improve the coverage and convergence speed according to the relationship among the angles of different sensors. By introducing the global optimal position to update the particle's position and defining the virtual force of uncovered grids for poor particles, Yao *et al.* [29] proposed a coverage control algorithm based on an improved gravitational search algorithm, which effectively improved the convergence speed. The work of [30] proposed a constrained artificial fish-swarm algorithm to optimize the sensor distribution by regarding the sensing centroid as artificial fish, which improved the coverage rate in the monitoring area. Utilizing discrete army ant search optimizer (AASO) in [31], the authors maximized the coverage of the target of DSNs. Fan *et al.* [32] defined the objective function about the ideal weighted coverage rate and searched the solution space by the quantum genetic algorithm. When the number of sensor nodes is given, the algorithm can obtain the maximum effective coverage rate of WMSNs. Zhu *et al.* [33] addressed the minimum-cost DSN deployment problem by using greedy heuristics, local search, and PSO.

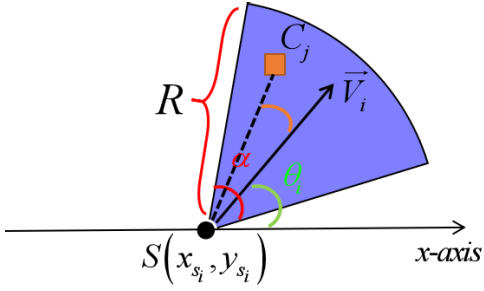

Fig. 3. Sensing model.

Coverage enhancement strategies based on geometry represented by VFA have the ability of quick convergence, however, it is difficult to set the optimal rotation angle of nodes at each time, resulting in an upper limit of coverage rate. In addition, VFA cannot achieve the best coverage effect in some cases even if the number of sensors is sufficient. For the coverage enhancement strategies based on SI, the no free lunch (NFL) theorem has logically proved that there is no optimization algorithm best suited for solving all the optimization problems [34]. The existing SI algorithm is easy to fall into the local optimal solution when solving the coverage problem of WMSNs. Therefore, the motivation of our research is to propose a novel SI algorithm, which can not only shows good competitiveness under the benchmark function, but also successfully solve the coverage problem of WMSNs.

The central contributions of our research are as follows.
1) The CEPWs is transformed into a constrained optimization problem, and its NP-hard is proved.
2) Inspired by the prey selection mechanism and information sharing mechanism in the process of army ants hunting, a novel AASO is first proposed in this article. The simulation results show that the optimizer has merits in terms of exploration and exploitation when compared with representative SI algorithms.
3) For coverage holes caused by random deployment, AASO is utilized to solve the coverage enhancement problem of WMSNs. Simulation experiments are performed and compared to existing approaches under different parameters, and the reasons for their performance differences in terms of coverage effect are revealed and discussed.

The article is organized as follows. In Section II, the coverage model of the WMSNs and the definition of the CEPW are introduced, while the proposed algorithm AASO and its utilization for CEPW are described in Section III. Simulation results and analysis of the AASO, IAPSO, improved ant lion optimizer (IALO), and VFA are given in Section IV. In Section V, we conclude the contribution of the article and highlight future research directions. Besides, the key parameters used in this article are listed in Table I.

TABLE I
KEY PARAMETERS USED IN THIS ARTICLE

| Parameters | Explanation |
|---|---|
| $S_i$ | The $i-th$ sensor node |
| $C_j$ | The centroid of $j-th$ grid |
| $P_{COVR}$ | The coverage rate of WMSNs |
| $D$ | The number of sensor nodes |
| $H$ | The area of monitoring region |
| $Ant_i$ | The $i-th$ army ant |
| $Prey_j$ | The $j-th$ prey |
| $T_{max}$ | The maximum number of iterations |
| $num_t^{aver}$ | The average number of army ants recruited by prey in the $t-th$ |
| $Prey_{for ant_i}^t$ | The attack direction and position of $Ant_i$ in the $t-th$ |
| $preynumber(t)$ | The number of prey in the $t-th$ |
| $Antbridge$ | The position of the ant bridge |

## II. COVERAGE MODEL

We consider a WMSN with $D$ multimedia sensor nodes in the monitoring area $L \times W$ and assume that all the nodes can achieve accurate positioning by using advanced positioning algorithms [35] and send the location information to the control center based on the preset routing protocol [10], [11]. Besides, this article assumes that there is no communication conflict, which can be guaranteed by the media access control (MAC) protocol based on time division multiple access (TDMA). We choose not to discuss them in this article, for the purpose of avoiding alleviating the focus of coverage enhancement to positioning and routing. In this section, we introduce the multimedia sensing model and prove the NP-hard of the CEPW.

### A. Multimedia Sensing Model

The sensing area of the multimedia sensor node is limited by the visual angle, as shown in Fig. 3. For any multimedia sensor $S_i$, which can be expressed as a 6-tuple $< (x_{s_i}, y_{s_i}), R, \vec{V_i}, \theta_i, \alpha >$, where $(x_{s_i}, y_{s_i})$ is the position of direction sensor $S_i$, $R$ is the sensing radius, $\vec{V_i}$ is the sensing direction, $\theta_i$ is the deviation angle between $\vec{V_i}$ and the $x$-axis, and $\alpha$ is the angle of view.

To calculate the coverage rate (COVR) of WMSNs, the monitoring area is divided into $M$ grids. A grid $C_j$ being sensed successfully by $S_i$ once (1) and (2) are both satisfied [29]

$$\text{dis}(S_i, C_j) = \|S_i, C_j\| \leq R \tag{1}$$

$$\overrightarrow{S_i C_j} \bullet \vec{V_i} \geq \|S_i, C_j\| \bullet \cos\left(\frac{\alpha}{2}\right) \tag{2}$$

where $\|S_i, C_j\|$ is the Euclidean distance between the centroid of grid $C_j$ and node $S_i$. A grid will be covered successfully once it is sensed by any node, which is expressed as $p_{c_j}$. The COVR of WMSNs be calculated by

$$P_{\text{COVR}} = \sum_{j=1}^{m} p_{c_j}/M. \tag{3}$$

Additionally, according to literature [16], assuming that $D$ multimedia sensor nodes are randomly deployed in the monitoring region with an area of $H$, and both the position of the node and its sensor direction follow the uniform distribution,

the initial COVR of the monitoring region being detected by $D$ multimedia sensor nodes is defined as

$$P_{\text{Ini}} = 1 - \left(1 - \frac{\alpha R^2}{2H}\right)^D. \quad (4)$$

Hence, when the COVR reaches at least $P_{\text{Ini}}$, the number of multimedia sensor nodes needing to be deployed is

$$D \geq \frac{\ln(1 - P_{\text{Ini}})}{\ln(2H - \alpha R^2) - \ln 2H}. \quad (5)$$

### B. Definition of the CEPW

Given a monitoring region $A$, a set of $D$ multimedia sensors $(S_1, S_2, \ldots, S_D)$ and its initial deviation angles $(\theta_1, \theta_2, \ldots, \theta_D)$, the CEPW is summarized as finding the optimal deviation angles $(\theta_1^*, \theta_2^*, \ldots, \theta_D^*)$ to satisfy the coverage requirement, which can be abstracted as a bound-constrained optimization problem

$$\max \left(P_{\text{COVR}}\left(\theta_1^*, \theta_2^*, \ldots, \theta_D^*\right)\right)$$
$$\text{s.t.} \quad 0 \leq \theta_i^* \leq 2\pi, \quad i = 1, 2, \ldots, D. \quad (6)$$

### C. Hardness of the CEPW

To prove the NP-hard of the CEPW, we define a special case of CEPW, namely CEPW*, and prove that it is NP-hard, which naturally induces the NP-hard of the original CEPW problem.

*Proof:* In CEPW*, given a set of $m$ grids to be covered; a set of $D$ multimedia sensors, each of which only has $p$ possible orientations, is randomly deployed in the monitoring area. Hence, CEPW* refers to drive the maximum number of grids to be covered, which can be treated as the classic maximum coverage problem, which is known to be NP-hard. As CEPW* is a special case of the CEPW, CWPW is also NP-hard.

## III. PROPOSED ALGORITHM

To solve the coverage enhancement of WMSNs, a novel SI technology AASO is first proposed inspired by the prey selection mechanism and information sharing mechanism of army ants in this section. Besides, we introduce the coverage enhancement strategy in WMSNs based on AASO.

### A. Biological Model

New world army ants live in species-rich assemblages through the Neotropics and are voracious predators of other ant general and arthropods. They are therefore an important and potentially informative group. There exist powerful information sharing mechanisms and prey selection mechanisms within the army ant population. Specifically, to increase the success of predation, the army ants have a strong response to the odor of prey, which triggered both increased localized recruitment and slowed the advancement of the raid as they targeted the odor [37]. There are many sources of species-specific odors cues that army ants could be detected, for example, cuticular hydrocarbons (CHCs) and colony odors. The odors of prey mainly include the following four.

*1) Nest Material Odors:* Nest material provides a source of gestalt colony odor, including high concentrations of CHCs, to the army ants without the presence of any adult prey.

*2) Living Ant Odors:* Living ants provide army ants with CHCs and alarm pheromones. In addition, clues of movement or vibration from living ants can also be effectively recognized by army ants.

*3) Dead Ant Odors:* Dead ants present the CHCs of the ants to the army ant raids without volatile alarm pheromones or movement cues.

*4) Alarm Odors:* Alarm odors refer to the volatile pheromones used in alarm and recruitment responses from living ants.

On the whole, the four odors are the potential position of prey. Besides, army ants react most strongly to the odor emitted by nest material odors and get the highest reward for predation, and their response to living ant odors, dead ant odors, and alarm odors is weakened in turn.

### B. Mathematical Model

The mathematical model of AASO is as follows, mainly including five operators.

*1) Army Ant and Prey Initialization:* In AASO, it is supposed that the position of the army ant is represented the problem's variables. Army ant can change position in 1-D, 2-D, 3-D, or hyper-dimensional space. The initial position of all army ants is randomly distributed in the search space, which can be expressed as

$$\text{Ant} = \left[\text{Ant}_{i,1}, \text{Ant}_{i,2}, \ldots, \text{Ant}_{i,D}\right], \quad i = 1, 2 \ldots N \quad (7)$$

where $N$ is the number of army ants and $D$ is the number of the variables. In the optimization problem, we have no idea about the position of the optimal solution. Therefore, the best four solutions obtained by initial positions of army ants are saved as the initial positions of the nest material, living ant, dead ant, and alarm odors, respectively, which can be expressed as

$$\text{Prey} = \left[\text{Prey}_{i,1}, \text{Prey}_{i,2}, \ldots, \text{Prey}_{i,D}\right], \quad i = 1, 2, 3, 4. \quad (8)$$

*2) Recruited by Prey:* Army ants have a strong response to the potential odors of prey, namely the odors have a high recruitment effect on army ants. However, considering the influence of weather, odor volatilization, and other factors, not every army ant can smell all odors from different kinds of prey. The number of army ants attracted by one of the prey in each iteration is described by Poisson distribution, which is suitable for expressing the number of random events per unit of time. In the design of AASO, Poisson distribution is used to describe the number of army ants recruited by one of the prey successfully in each iteration. Specifically, assuming that the number of army ants attracted by one of the prey in each iteration follows Poisson distribution, that is, the probability that there are $k$ army ants in the population Ant attracted by one of the prey in the $t$th is defined as

$$P(k \text{ ants in Ant}) = \frac{(\text{num}_t^{\text{aver}})^k}{k!} e^{-\text{num}_t^{\text{aver}}}, \quad k = 0, 1, \ldots N \quad (9)$$

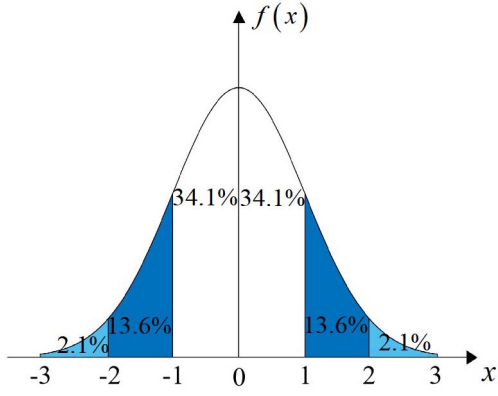

Fig. 4. Standard normal distribution.

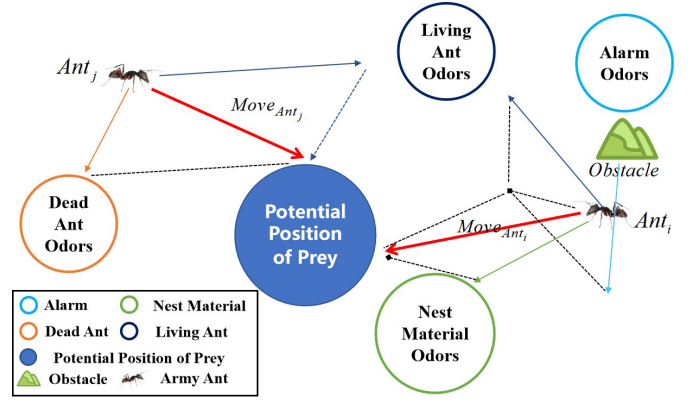

Fig. 6. Position update of army ants.

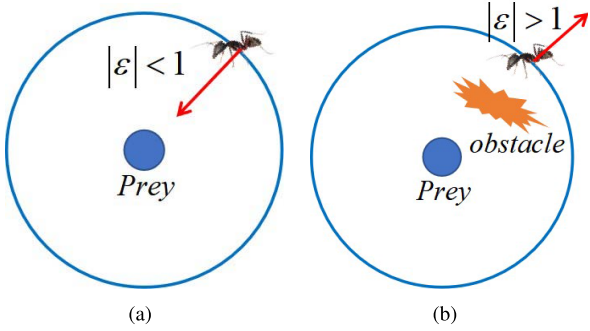

Fig. 5. Recruited by prey.

where $\text{num}_t^{\text{aver}}$ is the average number of army ants recruited by prey in the $t$th. Prey will have better fitness value and higher recruitment rate with the increase of iteration, and $\text{num}_t^{\text{aver}}$ can be expressed as

$$\text{num}_t^{\text{aver}} = \text{num}_{\text{ini}} + (N - \text{num}_{\text{ini}}) \times \frac{t}{T_{\max}} \quad (10)$$

where $T_{\max}$ is the maximum number of iterations and $\text{num}_{\text{ini}}$ is the average number of army ants initially recruited by prey. Based on the probability described in (9), the exact number of army ants recruited by one of the prey is selected by Roulette. In addition, the index of army ants can be randomly selected from the whole population to enhance the exploration of AASO.

Army ants will randomly scatter around the prey in the Gaussian distribution once they are recruited by the prey in the hyper-dimensional space. For example, if $\text{Ant}_i$ is recruited by $\text{Prey}_j$ in the $t$th, its position is calculated as

$$\text{Position}_{i-j}^t = \text{Prey}_j^t + \left(\text{Prey}_j^t - \text{Ant}_i^t\right) \odot [N(0,1)]^{1\times D} \quad (11)$$

where $\odot$ is the Hadamard product operator, and $[N(0,1)]^{1\times D}$ is a vector of $1 \times D$ dimension, in which the elements are random numbers with standard Gaussian distribution.

Given any number $\varepsilon$ following standard Gaussian distribution, as shown in Figs. 4 and 5(a), the probability of $|\varepsilon| < 1$ is 68.2%, which reflects the continuous exploit for the prey. Fig. 5(b) also shows that $|\varepsilon| > 1$ forces the army ant to diverge from the prey to hopefully explore a better prey. Generally speaking, the design of $|\varepsilon| > 1$ can also be regarded as the effect of obstacles in nature appearing in the hunting path, which prevents army ants from quickly and conveniently approaching prey.

*3) Attack Prey:* In each iteration, army ants will determine the attack direction and position based on the recruitment of prey. For example, the position update of the $\text{Ant}_i$ is expressed as

$$\text{Ant}_i^{t+1} = \text{Ant}_i^t + a \times r \times \left(\text{Prey}_{\text{forant}_i}^t - \text{Ant}_i^t\right) \quad (12)$$

$$\text{Prey}_{\text{forant}_i}^t = \sum_{j=1}^{\text{num}_{\text{forant}_i}^t} \text{Position}_{i-j}^t / \text{num}_{\text{forant}_i}^t \quad (13)$$

where $\text{Prey}_{\text{forant}_i}^t$ is the attack direction and position of $\text{Ant}_i$ in the $t$th, which is determined by the recruitment of different kinds of prey, described as $\text{Position}_{i-j}^t$ in (11), $r$ is the random number $(0,1]$, $\text{num}_{\text{forant}_i}^t$ is the number of prey recruiting $\text{Ant}_i$ in the $t$th, and $a$ is the attack coefficient, usually a value of 2. Obviously, the range of $\text{num}_{\text{forant}_i}^t$ is $[0,4]$.

Fig. 6 shows how $\text{Ant}_i$ and $\text{Ant}_j$ update their position to the prey in 2-D search space. For example, $\text{Ant}_i$ is recruited by alarm odors, living ant odors, and nest material odors, where obstacle forcing the $\text{Ant}_i$ diverge from the alarm odors ($|\varepsilon| < 1$) to explore more fitness prey.

There is a powerful information-sharing mechanism within the army ant population, if no prey recruiting $\text{Ant}_i$ in the $t$th, $\text{Ant}_i$ will randomly follow the two companions in the whole population to update the position, which can be expressed as

$$\text{Ant}_i^{t+1} = \sum_{k=1}^{2} \left(\text{Ant}_k^t + [\text{Cauchy}(0,1)]^{1\times D}\right)/2 \quad (14)$$

where $[\text{Cauchy}(0,1)]^{1\times D}$ is a vector of $1 \times D$ dimension, in which the elements are random numbers with standard Cauchy distribution. Compared with the Gaussian distribution, the Cauchy distribution has better perturbation, which helps AASO explore the local space.

*4) Update Prey:* After each iteration, we save the prey number $(t)$ best solutions obtained so far and oblige the army ants to update their positions according to the position of the prey number $(t)$ best solutions, where prey number $(t)$ is the number of prey in the $t$th. With the raids of army ants, the number of prey continues to decline, which is expressed as

$$\text{preynumber}(t) = \text{round}\left(4 - 4 \times \frac{t-1}{T_{\max}}\right). \quad (15)$$

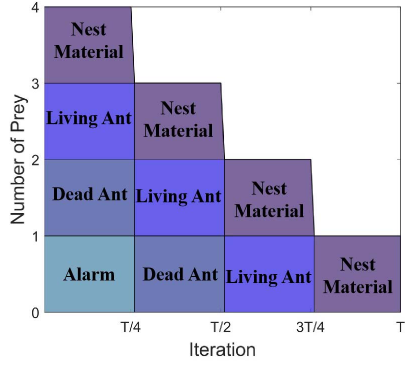

Fig. 7. Number of prey is decreased over the course of iterations.

Fig. 7 shows how the number of prey is decreased adaptively over the course of iterations. In AASO, the main consideration here is that the position updating of army ants with respect to different positions in the search space may degrade the exploitation of the best promising solutions increases.

*5) Build Ant Bridge:* Ant bridge is a way of mutual assistance among army ants; the army ants bite together one after another to form an ant bridge when they encounter a ravine, so that the entire population can pass quickly without considering the death of individuals. Aiming at the optimization problem, an ant bridge is built to mutate the positions of army ants so as to jump out of the local optimal. Ant bridge operator is an opening design, and different ant bridges can be built to help army ants search globally for different optimization problems. In this article, the design criteria of the ant bridge operator are as follows.

The algorithm may fall into the local optimal solution once the position of the nest material does not change in multiple iterations, the latter $N/2$ population with poor fitness will build an ant bridge to explore new positions. The position of the ant bridge is shown in (16), as shown at the bottom of the next page, where $\omega_k$ is the weight of army ants, which can be expressed as $\omega_k = F_k / \sum_{k=N/2+1}^{N} F_k$, where $F_k = 1/f(\text{Ant}_k)$, $f(\bullet)$ is the objective function. Army ants take the ant bridge as a benchmark and select a dimension for mutation arbitrarily. For example, the position after passing through the ant bridge is calculated in (17), as shown at the bottom of the next page, once the $j$th mutation dimension is selected by $\text{Ant}_k$. where $k$ is a random number in the range $(0, 1]$. The mutated position is compared with the position of the previous generation and the better one is retained for the next generation. The flowchart of the AASO is illustrated in Fig. 8.

### C. Parameter Analysis

The average number of army ants initially recruited by prey is an important parameter of AASO. Specifically, prey represents the best position searched by AASO, and army ants will determine the attack direction and position based on the recruitment of prey. Therefore, the value is quite large and may introduce large perturbations, which may have better exploration ability. Otherwise, AASO will have better exploitation ability. Therefore, how to set the appropriate value to make a good compromise between exploration and exploitation becomes the key problem. In this section, as shown in Table II, different values are set and recorded results through rigorous experimentation on six CEC 2017 benchmark functions (see Table IV for CEC 2017 benchmark functions). In the present work, $N/2$ provides satisfactory performance for most of the benchmark functions, and therefore $N/2$ is chosen as the initial value. It is worth explaining that the initial value can be adjusted for different optimization problems.

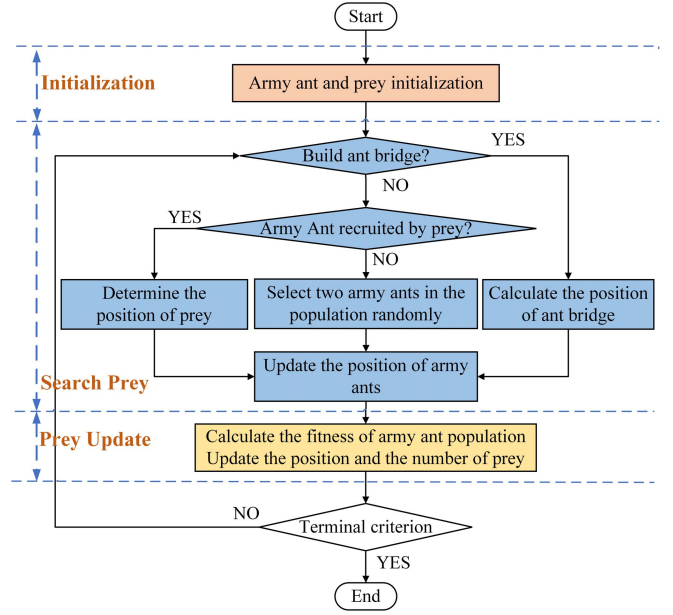

Fig. 8. Flowchart of AASO.

### D. Computational Complexity of AASO

As explained in Section III-B, AASO mainly includes five operators: army ant and prey initialization, recruit by prey, attack prey, update prey, and build ant bridge. The time complexity of each operators is, respectively, $O(N*D) + O(N^2)$, $O(N)$, $O(N*D)$, $O(N^2)$, and $O(N*D)$. Hence, the computational complexity of AASO is $O(N*D) + O(N^2) + O(T_{\max}*(N*D+N^2))$, which can be simplified to $O(T_{\max}*(N*\max\{N,D\}))$. It is obvious that the complexity of AASO only depends on the number of army ants size $N$, the dimension of the problem $D$, and the maximum number of iterations $T_{\max}$.

### E. Coverage Enhancement for WMSNs With AASO

For the coverage enhancement problem of finding the optimal deviation angles, the corresponding relationship between AASO and coverage enhancement for WMSNs based on AASO is shown in Table III. AASO is used to solve the bound constrained optimization problem of maximizing the coverage rate of WMSNs. The main steps are as follows.

TABLE II
Effect of the Average Number of Army Ants Recruited by Prey on the Performance of AASO on Six CEC 2017 Benchmark Functions

| Function | | 2N/3 | N/2 | N/3 | N/6 | Functions | | 2N/3 | N/2 | N/3 | N/6 |
|---|---|---|---|---|---|---|---|---|---|---|---|
| F5 | Best | 5.92E+02 | 5.51E+02 | 5.43E+02 | 5.41E+02 | F10 | Best | 3.17E+03 | 3.05E+03 | 3.47E+03 | 3.14E+03 |
| | Mean | 6.08E+02 | 5.96E+02 | 6.01E+02 | 5.79E+02 | | Mean | 4.43E+03 | 4.46E+03 | 4.52E+03 | 4.72E+03 |
| F13 | Best | 2.52E+03 | 2.77E+03 | 3.46E+03 | 4.89E+03 | F17 | Best | 1.79E+03 | 2.00E+03 | 2.12E+03 | 1.77E+03 |
| | Mean | 1.82E+04 | 1.94E+04 | 2.30E+04 | 2.43E+04 | | Mean | 2.21E+03 | 2.18E+03 | 2.19E+03 | 2.07E+03 |
| F21 | Best | 2.37E+03 | 2.36E+03 | 2.34E+03 | 2.34E+03 | F29 | Best | 3.41E+03 | 3.54E+03 | 3.38E+03 | 3.40E+03 |
| | Mean | 2.40E+03 | 2.39E+03 | 2.40E+03 | 2.39E+03 | | Mean | 3.94E+03 | 3.82E+03 | 3.84E+03 | 3.71E+03 |

TABLE III
Corresponding Relationship Between AASO and Coverage Enhancement for WMSNs Based on AASO

| AASO | Coverage Enhancement for WMSNs Based on AASO |
|---|---|
| Army ant | The set of deployment scheme |
| Prey | The optimal deployment schemes |
| Population size | Number of deployment schemes |
| Dimension of army ant | Number of multimedia sensor nodes |
| Position of army ant | deviation angle |
| Best fitness | Maximum coverage rate |

*Step 1:* The control center collects the information of multimedia sensor nodes $S = \{S_1, S_2, \ldots, S_D\}$.

*Step 2:* Initialize the set of deployment schemes. This procedure is corresponding to the step of army ant and prey initialization in AASO, which is described in (7) and (8). The set of deployment schemes is expressed as

$$\Theta = [\theta_{i,1}, \theta_{i,2}, \ldots, \theta_{i,k}, \ldots, \theta_{i,D}], \quad i = 1, 2, \ldots, N \quad (18)$$

where each deployment scheme contains $D$ deviation angles. For example, $\theta_{i,k}$ represents the deviation angle of multimedia sensor node $S_k$ in the $i$th deployment scheme.

Considering that AASO is utilized to solve the minimum problem, the CEPW belongs to the maximum optimization problem, and therefore the fitness function of CEPW is transformed as

$$F_{\text{CEPW}} = 1/P_{\text{COVR}} = M / \sum_{j=1}^{m} p_{c_j}. \quad (19)$$

The better four fitness obtained by the set of initial deployment schemes are saved as the initial optimal deployment schemes, which can be expressed as

$$\Theta^{\text{prey}} = \left[\theta_{j,1}^{\text{prey}}, \theta_{j,2}^{\text{prey}}, \ldots, \theta_{j,D}^{\text{prey}}\right], \quad j = 1, 2, 3, 4. \quad (20)$$

It is worth noting that the initial optimal deployment schemes are similar to the initial potential position of prey in AASO.

*Step 3:* Update the set of deployment schemes. This procedure is corresponding to the step of recruited by prey, attack prey, and build ant bridge in AASO. The set of deployment schemes and the optimal deployment schemes are regarded as army ant and prey, respectively. In other words, the set of deployment schemes will update the deviation angles based on the optimal deployment schemes. More specifically, the set of deployment schemes is updated based on (9)–(14) and (16)–(17). Equations (9)–(14) help each node explore more deviation angles near the optimal deployment schemes. Equations (16) and (17) are helpful for a single node to explore local sensing blind areas and improve the local COVR.

*Step 4:* Update the optimal deployment schemes. This procedure is corresponding to the step of update prey in AASO. Specifically, the number of the optimal deployment schemes decreases gradually with the increase of iterations, which is expressed as

$$\Theta_{\text{preynumber}}(t) = \text{round}\left(4 - 4 \times \frac{t-1}{T_{\max}}\right) \quad (21)$$

where $t$ and $T_{\max}$ are, respectively, the current iteration round and the maximum iteration round.

*Step 5:* Terminate criterion controlling. The coverage enhancement for the WMSN process is terminated after a fixed number of iterations. Repeat steps 2 and 3 until the termination criterion is satisfied; otherwise, proceed to step 6.

*Step 6:* Calculate the final COVR based on the optimal deviation angle. The control center sends the deviation angle information to each node.

Algorithm 1 shows the pseudocode of coverage enhancement for WMSNs based on AASO.

## IV. Experimental Results

In this section, two experiments are performed with MATLAB 2016a on a computer with a 3.11-GHz frequency and 16-GB memory to evaluate the performance of our proposed algorithm AASO.

### A. AASO for CEC2017 Benchmark Functions

AASO is compared with representative SI algorithms including PSO, ant lion optimizer (ALO) [23], and salp

$$\text{Antbridge} = \left[\sum_{k=\frac{N}{2}+1}^{N} \omega_k \text{Ant}_{k,1}, \sum_{k=\frac{N}{2}+1}^{N} \omega_k \text{Ant}_{k,2}, \ldots, \sum_{k=\frac{N}{2}+1}^{N} \omega_k \text{Ant}_{k,D}\right] \quad (16)$$

$$\text{Ant}_k = \left(\text{Ant}_{k,1}, \text{Ant}_{k,2}, \ldots, 2k \times \text{Antbridge}_j - \text{Ant}_{k,j}, \ldots, \text{Ant}_{k,D}\right) \quad (17)$$

TABLE IV
DEFINITION OF CEC 2017 BENCHMARK

| No. | Types | Name | Optimum |
|---|---|---|---|
| F1 | Unimodal functions | Shifted and Rotated Bent Cigar Functions | 100 |
| F2 | | Shifted and Rotated Sum of Different Power Function | 200 |
| F3 | | Shifted and Rotated Zakharov Function | 300 |
| F4 | Simple Multimodal functions | Shifted and Rotated Rosenbrock's Functions | 400 |
| F5 | | Shifted and Rotated Rastrigin's Function | 500 |
| F6 | | Shifted and Rotated Expanded Scaffer's $F6$ Function | 600 |
| F7 | | Shifted and Rotated Lunacek Bi_Rastrigin Function | 700 |
| F8 | | Shifted and Rotated Non_Continuous Rastrigin's Function | 800 |
| F9 | | Shifted and Rotated Levy Function | 900 |
| F10 | | Shifted and Rotated Schwefel's Function | 1000 |
| F11 | Hybrid functions | Hybrid Function 1 (N=3) | 1100 |
| F12 | | Hybrid Function 2 (N=3) | 1200 |
| F13 | | Hybrid Function 3 (N=3) | 1300 |
| F14 | | Hybrid Function 4 (N=4) | 1400 |
| F15 | | Hybrid Function 5 (N=4) | 1500 |
| F16 | | Hybrid Function 6 (N=4) | 1600 |
| F17 | | Hybrid Function 6 (N=5) | 1700 |
| F18 | | Hybrid Function 6 (N=5) | 1800 |
| F19 | | Hybrid Function 6 (N=6) | 1900 |
| F20 | | Hybrid Function 6 (N=6) | 2000 |
| F21 | Composition functions | Composition Function 1 (N=3) | 2100 |
| F22 | | Composition Function 2 (N=3) | 2200 |
| F23 | | Composition Function 3 (N=4) | 2300 |
| F24 | | Composition Function 4 (N=4) | 2400 |
| F25 | | Composition Function 5 (N=5) | 2500 |
| F26 | | Composition Function 6 (N=5) | 2600 |
| F27 | | Composition Function 7 (N=6) | 2700 |
| F28 | | Composition Function 8 (N=6) | 2800 |
| F29 | | Composition Function 9 (N=3) | 2900 |
| F30 | | Composition Function 10 (N=3) | 3000 |

Search Range: $[-100, 100]^D$

---

**Algorithm 1** Coverage Enhancement for WMSNs Based on AASO.
**Require:** Multimedia sensor nodes: $S = \{S_1, S_2, \ldots, S_D\}$; Predefined number of deployment schemes: $N$; Maximum iterations: $T_{\max}$.
**Ensure:** The deviation angle of sensors; Coverage rate.
1: Initialize the set of deployment scheme: $\Theta_{i,j}, \forall i, j, 1 \leq i \leq N, 1 \leq j \leq D$
2: **for** $i = 1 \rightarrow N$ **do**
   Calculate the fitness of $\theta_i$ using (19).
3: **end for**
4: Generate initial optimal deployment schemes.
5: **for** $t = 1 \rightarrow T_{max}$ **do**
6:   **for** $i = 1 \rightarrow N$ **do**
     Update the deviation angle information of $\theta_i$.
     Calculate the fitness of $\theta_i$ using (19).
7:   **end for**
8:   Update the optimal deployment schemes.
9: **end for**
10: Acquire the optimal deviation angle of sensors and coverage rate $P_{COVR}$.

---

swarm algorithm (SSA) [22] on the CEC 2017 benchmark suite, which includes three unimodal functions, seven simple multimodal functions, ten hybrid functions, and ten composition functions, as shown in Table IV. The population size, maximum iterations, and problem dimensions are, respectively, set to 30, 1000, and 30. For the statistical analysis, each benchmark is carried out for 50 independent runs to minimize the statistical error of the results. The relevant parameters in PSO, ALO, and SSA adopt the values set in the original algorithm.

Table V shows the comparison results of AASO and other SI algorithms on the CEC 2017 benchmark suite, the best performing values in each evaluation criterion are bold. AASO obtains the best mean value among 22 functions (F2–F3, F5–F17, F19, F21, F23, F27–F30), and SSA performed the optimal value in two (F22, F26). AASO and SSA show similar performance in three functions (F4, F24, F25), which are better than the other two algorithms. In addition, ALO and PSO achieved, respectively, the optimal value in two (F1, F18) and one (F20) benchmark functions. From the perspective of standard deviation, AASO obtained the 23 best results, and SSA, ALO, and PSO each has an optimal standard deviation on two test functions (F4, F23), three test functions (F1, F18, F28), and two test functions (F10, F22). Comparing the best results in 20 independent runs, it can be seen that AASO still has large merit. The experimental results also confirm the NFL theorem, not all evolutionary algorithms can perform optimally on all optimization problems. Also, Fig. 9 shows the convergence curves of the four algorithms on 12 CEC 2017 benchmark functions. Intuitively, AASO shows good convergence when compared with other algorithms.

The reasons provide very competitive results of AASO on the CEC 2017 benchmark suite and are analyzed in detail as follows. In AASO, a new direction for movement of the $i$th army ants is obtained, as shown in Fig. 8, in three ways: attacking prey, following two companions randomly, and building an ant bridge. Attacking prey and the decreasing number of prey reflect the continuous exploit for the optimal solution by AASO. In (14), the design of normal distribution also considers the counterattack when $|\varepsilon| > 1$ in Fig. 5, forcing army ants to be more difficult to approach the prey, which reflects the exploration of AASO. When army ants randomly follow two companions in the whole population to update the position, the Cauchy distribution provides local exploration ability of AASO due to a wider distribution range and greater variation compared with normal distribution. Ant bridge enhances the population diversity. In the early iteration, there are great differences among individuals, which can help ants explore more search spaces. In the later iteration, an ant bridge can still ensure the diversity of the population. In summary, AASO makes a good compromise between exploration and exploitation.

TABLE V
SIMULATION RESULTS OF AASO VERSUS OTHER SI ON CEC 2017 WITH 30 DIMENSIONS

| Functions | | PSO | ALO | SSA | AASO | Functions | | PSO | ALO | SSA | AASO |
|---|---|---|---|---|---|---|---|---|---|---|---|
| F1 | Best | 4.13E+09 | 2.14E+02 | 1.33E+02 | **1.13E+02** | F2 | Best | 8.32E+21 | 2.03E+10 | 2.03E+12 | **1.63E+06** |
| | Mean | 1.22E+10 | **2.21E+03** | 4.95E+03 | 3.61E+03 | | Mean | 1.53E+41 | 5.47E+15 | 4.26E+17 | **4.93E+10** |
| | Std | 6.98E+09 | **1.42E+03** | 5.10E+03 | 4.60E+03 | | Std | 6.65E+41 | 2.34E+16 | 8.50E+17 | **9.82E+10** |
| F3 | Best | 6.68E+04 | 4.54E+04 | 1.25E+04 | **6.72E+03** | F4 | Best | 6.50E+02 | **4.63E+02** | 4.75E+02 | 4.65E+02 |
| | Mean | 1.25E+05 | 1.21E+05 | 3.33E+04 | **2.06E+04** | | Mean | 2.19E+03 | 5.11E+02 | **5.02E+02** | **5.02E+02** |
| | Std | 5.24E+04 | 4.57E+04 | 1.44E+04 | **8.51E+03** | | Std | 1.35E+03 | 2.54E+01 | **1.80E+01** | 2.33E+01 |
| F5 | Best | 6.02E+02 | 6.20E+02 | 6.07E+02 | **5.51E+02** | F6 | Best | 6.15E+02 | 6.29E+02 | 6.28E+02 | **6.02E+02** |
| | Mean | 6.65E+02 | 6.72E+02 | 6.64E+02 | **5.96E+02** | | Mean | 6.23E+02 | 6.47E+02 | 6.48E+02 | **6.07E+02** |
| | Std | 3.22E+01 | 3.07E+01 | 4.19E+01 | **2.58E+01** | | Std | 7.56E+00 | 1.00E+01 | 9.72E+00 | **4.93E+00** |
| F7 | Best | 8.34E+02 | 8.86E+02 | 8.40E+02 | **7.98E+02** | F8 | Best | 8.99E+02 | 8.92E+02 | 8.78E+02 | **8.55E+02** |
| | Mean | 1.03E+03 | 1.00E+03 | 9.17E+02 | **8.68E+02** | | Mean | 9.60E+02 | 9.46E+02 | 9.49E+02 | **8.93E+02** |
| | Std | 1.54E+02 | 8.70E+01 | 5.40E+01 | **4.22E+01** | | Std | 3.98E+01 | 3.49E+01 | 5.49E+01 | **2.13E+01** |
| F9 | Best | 2.20E+03 | 1.67E+03 | 3.15E+03 | **1.06E+03** | F10 | Best | 4.43E+03 | 4.38E+03 | 3.85E+03 | **3.05E+03** |
| | Mean | 5.52E+03 | 3.70E+03 | 5.24E+03 | **1.99E+03** | | Mean | 5.43E+03 | 5.50E+03 | 5.00E+03 | **4.46E+03** |
| | Std | 2.46E+03 | 1.23E+03 | 1.18E+03 | **8.29E+02** | | Std | **5.72E+02** | 6.31E+02 | 8.12E+02 | 9.36E+02 |
| F11 | Best | 1.32E+03 | 1.17E+03 | 1.24E+03 | **1.15E+03** | F12 | Best | 4.24E+07 | 9.47E+05 | 1.44E+06 | **5.88E+04** |
| | Mean | 1.87E+03 | 1.36E+03 | 1.33E+03 | **1.24E+03** | | Mean | 1.65E+09 | 1.08E+07 | 1.18E+07 | **1.15E+06** |
| | Std | 1.06E+03 | 8.54E+01 | 7.26E+01 | **5.92E+01** | | Std | 1.88E+09 | 7.03E+06 | 1.17E+07 | **1.10E+06** |
| F13 | Best | 2.08E+05 | 6.01E+04 | 3.75E+04 | **2.77E+03** | F14 | Best | 1.36E+04 | **4.55E+03** | 6.37E+03 | 4.69E+03 |
| | Mean | 4.92E+08 | 1.41E+05 | 1.37E+05 | **1.94E+04** | | Mean | 7.88E+05 | 6.19E+04 | 5.38E+04 | **5.05E+04** |
| | Std | 9.05E+08 | 7.46E+04 | 1.04E+05 | **1.68E+04** | | Std | 1.58E+06 | 5.58E+04 | 4.20E+04 | **3.63E+04** |
| F15 | Best | 1.01E+04 | 1.97E+04 | 1.62E+04 | **1.79E+03** | F16 | Best | 2.52E+03 | 2.42E+03 | 2.25E+03 | **2.20E+03** |
| | Mean | 8.82E+07 | 5.21E+04 | 7.69E+04 | **1.22E+04** | | Mean | 3.06E+03 | 3.11E+03 | 2.93E+03 | **2.49E+03** |
| | Std | 3.94E+08 | 2.99E+04 | 6.14E+04 | **1.08E+04** | | Std | 3.86E+02 | 4.24E+02 | 3.82E+02 | **1.81E+02** |
| F17 | Best | **1.78E+03** | 2.10E+03 | 1.91E+03 | 2.00E+03 | F18 | Best | 2.68E+05 | 5.45E+04 | **3.81E+04** | 1.27E+05 |
| | Mean | 2.35E+03 | 2.42E+03 | 2.29E+03 | **2.18E+03** | | Mean | 3.09E+06 | **7.01E+05** | 1.51E+06 | 9.52E+05 |
| | Std | 2.81E+02 | 2.32E+02 | 2.70E+02 | **1.54E+02** | | Std | 6.47E+06 | **6.15E+05** | 1.99E+06 | 1.27E+06 |
| F19 | Best | 1.46E+04 | 4.05E+04 | 3.67E+04 | **1.96E+03** | F20 | Best | **2.14E+03** | 2.22E+03 | 2.36E+03 | 2.24E+03 |
| | Mean | 8.09E+07 | 1.96E+06 | 2.36E+06 | **7.68E+03** | | Mean | **2.47E+03** | 2.58E+03 | 2.62E+03 | 2.48E+03 |
| | Std | 3.06E+08 | 1.45E+06 | 1.54E+06 | **7.09E+03** | | Std | 1.96E+02 | 2.32E+02 | 2.03E+02 | **1.23E+02** |
| F21 | Best | 2.39E+03 | 2.38E+03 | 2.39E+03 | **2.36E+03** | F22 | Best | 3.01E+03 | 2.30E+03 | **2.30E+03** | **2.30E+03** |
| | Mean | 2.47E+03 | 2.46E+03 | 2.44E+03 | **2.39E+03** | | Mean | 5.90E+03 | 4.55E+03 | **4.32E+03** | 5.52E+03 |
| | Std | 3.73E+01 | 3.92E+01 | 3.35E+01 | **2.12E+01** | | Std | **1.30E+03** | 2.33E+03 | 1.97E+03 | 1.48E+03 |
| F23 | Best | 2.79E+03 | 2.83E+03 | **2.70E+03** | 2.71E+03 | F24 | Best | 3.00E+03 | 3.03E+03 | **2.88E+03** | **2.88E+03** |
| | Mean | 2.99E+03 | 2.93E+03 | 2.78E+03 | **2.77E+03** | | Mean | 3.16E+03 | 3.14E+03 | **2.93E+03** | **2.93E+03** |
| | Std | 9.26E+01 | 5.98E+01 | **4.49E+01** | 4.93E+01 | | Std | 1.03E+02 | 7.83E+01 | 4.10E+01 | **3.66E+01** |
| F25 | Best | 2.92E+03 | **2.89E+03** | **2.89E+03** | **2.89E+03** | F26 | Best | 4.47E+03 | **2.80E+03** | **2.80E+03** | **2.80E+03** |
| | Mean | 3.20E+03 | 2.93E+03 | **2.91E+03** | **2.91E+03** | | Mean | 6.99E+03 | 5.92E+03 | **4.76E+03** | 4.78E+03 |
| | Std | 2.64E+02 | **1.94E+01** | 3.38E+01 | 2.23E+01 | | Std | 1.34E+03 | 2.02E+03 | **9.50E+02** | 1.08E+03 |
| F27 | Best | 3.29E+03 | 3.30E+03 | **3.21E+03** | **3.21E+03** | F28 | Best | 3.52E+03 | **3.21E+03** | 3.23E+03 | **3.20E+03** |
| | Mean | 3.40E+03 | 3.43E+03 | 3.26E+03 | **3.24E+03** | | Mean | 4.69E+03 | 3.26E+03 | 3.28E+03 | **3.23E+03** |
| | Std | 9.78E+01 | 8.16E+01 | 2.95E+01 | **1.30E+01** | | Std | 1.10E+03 | **2.36E+01** | 3.64E+01 | 2.41E+01 |
| F29 | Best | 3.64E+03 | 4.22E+03 | 3.85E+03 | **3.54E+03** | F30 | Best | 1.62E+05 | 4.92E+05 | 6.06E+05 | **5.46E+03** |
| | Mean | 4.22E+03 | 4.68E+03 | 4.18E+03 | **3.82E+03** | | Mean | 8.80E+06 | 5.38E+06 | 4.97E+06 | **1.13E+04** |
| | Std | 3.75E+02 | 2.90E+02 | 1.98E+02 | **1.97E+02** | | Std | 1.08E+07 | 4.06E+06 | 3.46E+06 | **5.05E+03** |

### B. AASO for Coverage Enhancement in WMSNs

To evaluate the performance of the final COVR in WMSNs, AASO is compared with IALO [25], IAPSO, and VFA, where VFA includes attraction force from uncovered grids and repulsion force between sensor nodes. The main parameters are shown in Table VI.

A total of 110 multimedia sensor nodes, with $R = 60$ m and $\theta = \pi/2$, are randomly deployed in the monitoring area $500 \times 500$ m. Fig. 10 shows the initial coverage effect with COVR of 68.32%. Fig. 11(a)–(d) shows the final coverage effect of considered four algorithms, where the final COVR of IALO, IAPSO, and VFA are, respectively, 81.82%, 81.55%, and 82.89%. The final COVR of AASO can reach 87.52%, which is 19.20% higher than the initial COVR, and 5.70%, 5.97%, and 4.63% higher than those of the other three algorithms, respectively. Coverage optimization can significantly reduce the cost of deployment, with at least 183 nodes needing to be deployed if the expected COVR is 87.52% according to (5), namely AASO-based coverage enhancement strategy can save nearly 73 nodes. Fig. 12 shows the plot of COVR of the four algorithms with the iterations, when the iteration reaches 20 rounds, the COVR of AASO can reach 83.07%, which is better than the final COVR of the other four algorithms.

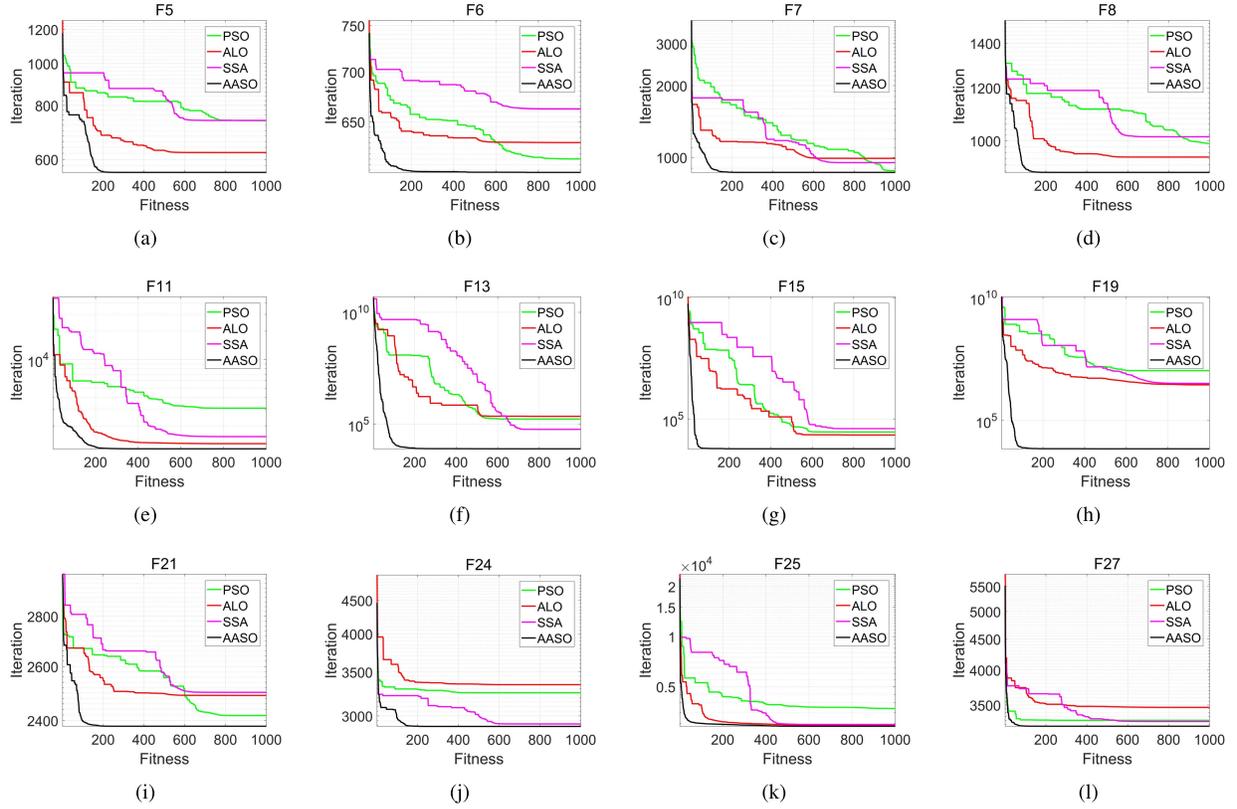

Fig. 9. Convergence curves of the four algorithms on 12 CEC 2017 benchmark functions.

TABLE VI
SIMULATION PARAMETERS

| Category | Parametes | Value |
| --- | --- | --- |
| General parameter | Monitoring Area | $500m \times 500m, 800m \times 800m, 1000m \times 1000m$ |
| | Number of multimedia sensor nodes | $50, 70, 90, 100, 110, 120, 130, 140$ |
| | Sensing radius | $40m, 60m, 80m, 100m$ |
| | Angle of view | $\pi/3, \pi/2,$ |
| | Discrete interval of grids | $3m, 5m, 7m$ |
| IALO | Number of ants and antlions | 50 |
| | Maximum number of iterations $T_{max}$ | 100 |
| | The range of antlions (Nmin,Nmax) | [30,50] |
| IAPSO | Number of particles | 50 |
| | Maximum number of iterations $T_{max}$ | 100 |
| | Acceleration cofficients ($c_1, c_2$) | 2.0 |
| | Maximum / Mimimum interia weight (Wmax, Wmin) | [0,1] |
| | Maximum Velocity $V_{max}$ | $2\pi$ |
| VFA | Maximum number of iterations $T_{max}$ | 100 |
| | Rotation angle | $\pi/90$ |
| AASO | Number of army ants | 50 |
| | Maximum number of iterations $T_{max}$ | 100 |
| | attack cofficient ($a$) | 2.0 |

To verify the reliability of AASO for CEPW, 50 experiments are performed independently with different parameters, as shown in Table VII. Intuitively, AASO always shows good performance in coverage effect compared with other algorithms under different deployment environments. The simulation results can firmly prove the reliability of AASO and verify the accuracy of the previous single experiment.

## TABLE VII
PERFORMANCE COMPARISON OF THE AVERAGE RESULTS OF 50 EXPERIMENTS

| Monitoring Area | Sensor Parameter | AASO | IALO | IAPSO | VFA |
|---|---|---|---|---|---|
| $500m \times 500m$ | $R=60m, D=100, \alpha=\pi/2$ | 82.84% | 77.56% | 78.72% | 76.31% |
| | $R=40m, D=100, \alpha=\pi/2$ | 48.08% | 44.52% | 46.02% | 44.52% |
| | $R=60m, D=100, \alpha=\pi/3$ | 65.97% | 61.72% | 59.98% | 58.32% |
| $800m \times 800m$ | $R=80m, D=120, \alpha=\pi/2$ | 75.47% | 72.21% | 72.59% | 74.13% |
| | $R=60m, D=120, \alpha=\pi/2$ | 48.38% | 45.26% | 45.69% | 46.21% |
| | $R=80m, D=120, \alpha=\pi/3$ | 55.79% | 51.89% | 50.71% | 51.02% |
| $1000m \times 1000m$ | $R=100m, D=140, \alpha=\pi/2$ | 80.98% | 73.72% | 73.46% | 77.21% |
| | $R=80m, D=140, \alpha=\pi/2$ | 61.72% | 58.53% | 59.28% | 59.62% |
| | $R=100m, D=140, \alpha=\pi/3$ | 62.39% | 59.15% | 58.76% | 58.59% |

## TABLE VIII
COMPARISON OF THE ALGORITHM COMPLEXITY AND RUNNING TIME OF THE FOUR ALGORITHMS

| Alogorthm | AASO | IALO | IAPSO | VFA |
|---|---|---|---|---|
| Complexity | $O(T_{max} * N * M * D)$ | $O(T_{max} * N * M * D)$ | $O(T_{max} * N * M * D)$ | $O(T_{max} * M * D)$ |
| Running time | 388.12s | 372.03s | 393.91s | 15.36s |

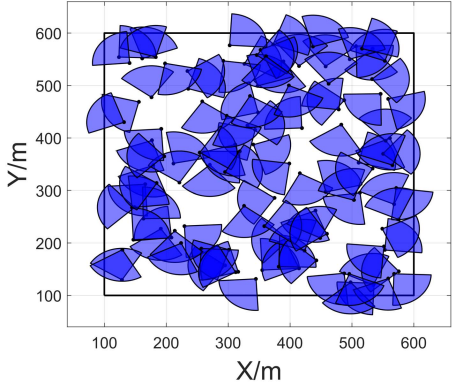

Fig. 10. Initial coverage effect (68.32%).

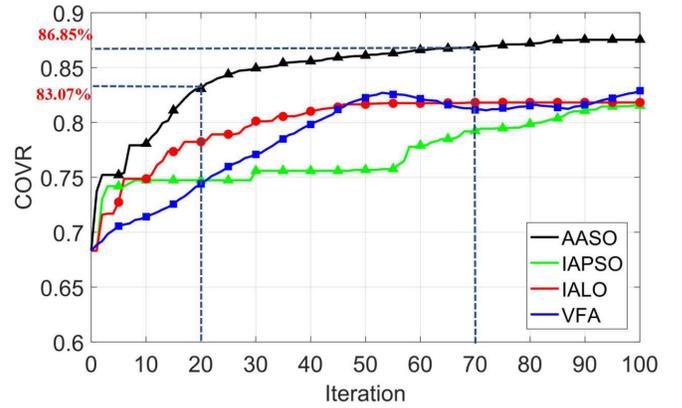

Fig. 12. Plot of COVR with the iterations of four algorithms.

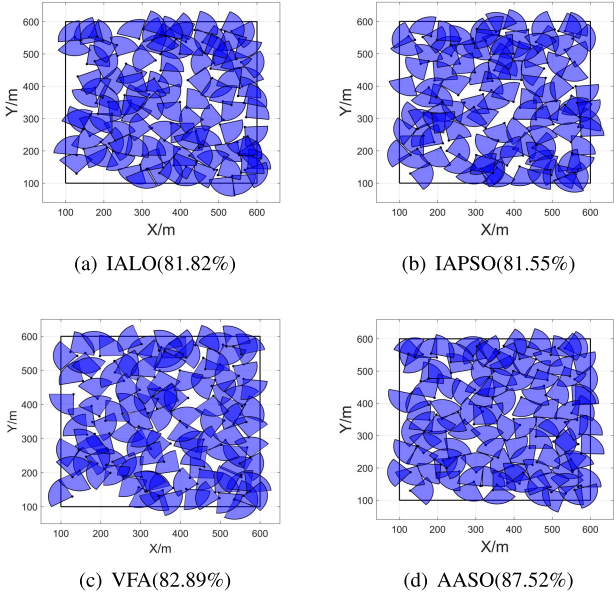

Fig. 11. (a)–(d) Final coverage effect of IALO, IAPSO, VFA, and AASO. (a) IALO (81.82%). (b) IAPSO (81.55%). (c) VFA (82.89%). (d) AASO (87.52%).

The reasons that lead to the performance differences of the final COVR of AASO, VFA, IAPSO, and IALO are analyzed in detail as follows. For VFA, as shown in Fig. 12, it is easy to cause oscillation at the end of the iteration. In view of the essence of VFA is to separate overlapping nodes and fill unmonitored areas according to the virtual force, it is noteworthy that the final COVR will be inevitably affected by the optimal rotation angle of nodes at each time. The optimal rotation angle is affected by the initial deployment and iterations, which is difficult to set. Although it has been adjusted repeatedly, the final COVR is still worse than AASO. For SI optimization technology, the optimization effect is closely related to the search space, dimension, and constraints of the problem. Given that the essence of ALO and its improved version is to imitate the hunting behavior of antlions in nature to search for the optimal solution, the dynamic update strategy of ant lion can accelerate the convergence speed of the algorithm. However, aiming at the coverage enhancement of WMSNs, the boundary shrinkage factor $I$ can easily lead to local optimization in a later iteration. AASO and IAPSO all initiate the optimization process by the movement of search agents in the search space, but the movement mechanism is entirely different. In IAPSO, the new direction for movement of the $i$th particle is affected by the cumulative effect of inertia weight, Pbest, Gbest, and velocity.

The adaptive weight in IAPSO provides more exploration when compared with classical PSO. However, it is difficult to set a suitable velocity threshold to balance the convergence speed and optimization ability. In AASO, Poisson distribution, which describes the number of army ants recruited by prey successfully in each iteration, determines the position and direction of the army ants' attack. More importantly, Cauchy distribution helps AASO explore more deployment strategies near the best particles in CEPW when compared to PSO. In addition, an ant bridge also allows a single node to explore more coverage holes. Therefore, AASO is more suitable for coverage enhancement in WMSNs.

In addition, the comparison of the algorithm complexity and running time of the four algorithms are shown in Table VIII. For a WMSN, with $D$ nodes and $M$ discrete grids, the algorithm complexity of AASO, IALO, IAPSO, and VFA are, respectively, $O(T_{\max} * N * M * D)$, $O(T_{\max} * N * M * D)$, $O(T_{\max} * N * M * D)$, and $O(T_{\max} * M * D)$. For the SI algorithm to solve the coverage enhancement problem, the number of population $N$, the maximum number of iterations $T_{\max}$, the number of nodes $D$, and the number of discrete grids $M$ are important components of complexity. The main reason for the difference in running time is the different design of operators in the algorithm, but it can be ignored when the number of population or problem dimensions is large enough. Besides, the algorithm complexity and running time of VFA are better than AASO. As an SI algorithm, considering that the essence of AASO is the competition among multiple deployment schemes to achieve better coverage, while VFA only updates one deployment scheme. Hence, it is affordable for AASO to optimize the coverage effect by sacrificing the complexity and running time.

## V. Conclusion

In this work, we propose a coverage enhancement strategy based on AASO in WMSNs. The predation behavior of army ants was the main inspiration for AASO, and three mathematical models were proposed to update the position of army ants. In addition, aiming at the coverage holes caused by random deployment, by transforming the coverage enhancement problem into a constrained optimization problem and proving its NP-hard, AASO is utilized to improve the coverage effect in WMSNs. Simulation results demonstrate that AASO shows improved performance in final COVR under different scenes when compared to three existing algorithms, and it also has superior robustness when the number of nodes, the sensing radius, and the angle of view change. Besides, we firmly believe that if different kinds of prey affect the updating of army ants' positions based on their fitness, better results will be achieved in some other optimization problems. However, the research of this article still has some limitations, for example, considering the challenges of sensing and communication ability due to the existence of obstacles, the sensing range, and the dropout rate are different in WMSNs. Hence, we will pay more attention to the application of AASO in WMSNs for remote monitoring in the future.

## References

[1] Y. Pan et al., "CDD: Coordinating data dissemination in heterogeneous IoT networks," *IEEE Commun. Mag.*, vol. 28, no. 6, pp. 84–89, Jun. 2020.

[2] Y. Lu and L. D. Xu, "Internet of Things (IoT) cybersecurity research: A review of current research topics," *IEEE Internet Things J.*, vol. 6, no. 2, pp. 2103–2115, Apr. 2019.

[3] C. Luo, Y. Cao, G. Xin, B. Wang, E. Lu, and H. Wang, "Three-dimensional coverage optimization of underwater nodes under multi-constraints combined with water flow," *IEEE Internet Things J.*, vol. 9, no. 3, pp. 2375–2389, Feb. 2022.

[4] G. Han, X. Yang, L. Liu, W. Zhang, and M. Guizani, "A disaster management-oriented path planning for mobile anchor node-based localization in wireless sensor networks," *IEEE Trans. Emerg. Topics Comput.*, vol. 8, no. 1, pp. 115–125, Jan. 2020.

[5] H. M. Jawad et al., "Accurate empirical path-loss model based on particle swarm optimization for wireless sensor networks in smart agriculture," *IEEE Sensors J.*, vol. 20, no. 1, pp. 552–561, Jan. 2020.

[6] C. Han, S. Zhang, B. Zhang, J. Zhou, and L. Sun, "A distributed image compression scheme for energy harvesting wireless multimedia sensor networks," in *Proc. 15th Int. Conf. Mobile Ad-Hoc Sensor Netw. (MSN)*, Dec. 2019, pp. 362–367.

[7] S. M. Aziz and D. M. Pham, "Energy efficient image transmission in wireless multimedia sensor networks," *IEEE Commun. Lett.*, vol. 17, no. 6, pp. 1084–1087, Jun. 2013.

[8] R. Komura and T. Nakano, "Inspection method of hyper spectral camera with spectrometer for forest observation," in *Proc. Int. Conf. Eng., Appl. Sci., Technol. (ICEAST)*, 2018, pp. 1–4.

[9] G. Baldoni, M. Melita, S. Micalizzi, C. Rametta, G. Schembra, and A. Vassallo, "A dynamic, plug-and-play and efficient video surveillance platform for smart cities," in *Proc. 14th IEEE Annu. Consum. Commun. Netw. Conf. (CCNC)*, Jan. 2017, pp. 611–612.

[10] X. Zhao, H. Zhu, S. Aleksic, and Q. Gao, "Energy-efficient routing protocol for wireless sensor networks based on improved grey wolf optimizer," *KSII Trans. Internet Inf. Syst.*, vol. 12, no. 6, pp. 2644–2657, Jun. 2018.

[11] P. Wu, F. Xiao, H. Huang, and R. Wang, "Load balance and trajectory design in multi-UAV aided large-scale wireless rechargeable networks," *IEEE Trans. Veh. Technol.*, vol. 69, no. 11, pp. 13756–13767, Jun. 2018.

[12] Y. Ye, L. Shi, X. Chu, R. Q. Hu, and G. Lu, "Resource allocation in backscatter-assisted wireless powered MEC networks with limited MEC computation capacity," *IEEE Trans. Wireless Commun.*, early access, Jun. 30, 2022, doi: 10.1109/TWC.2022.3185825.

[13] L. Zhang, T.-T. Liu, F.-Q. Wen, L. Hu, C. Hei, and K. Wang, "Differential evolution based regional coverage-enhancing algorithm for directional 3D wireless sensor networks," *IEEE Access*, vol. 7, pp. 93690–93700, 2019.

[14] T. O. Olasupo and C. E. Otero, "A framework for optimizing the deployment of wireless sensor networks," *IEEE Trans. Netw. Service Manag.*, vol. 15, no. 3, pp. 1105–1118, Sep. 2018.

[15] R. Elhabyan, W. Shi, and M. St-Hilaire, "Coverage protocols for wireless sensor networks: Review and future directions," *J. Commun. Netw.*, vol. 21, no. 1, pp. 45–60, Feb. 2019.

[16] D. Tao, H. D. Ma, and L. Liu, "A virtual potential field based coverage enhancing algorithm for directional sensor networks," *J. Software.*, vol. 18, pp. 1152–1163, May 2007.

[17] Z. Jing and Z. Jian-Chao, "A virtual centripetal force-based coverage-enhancing algorithm for wireless multimedia sensor networks," *IEEE Sensors J.*, vol. 10, no. 8, pp. 1328–1334, Aug. 2010.

[18] H. Ma, X. Zhang, and A. Ming, "A coverage-enhancing method for 3D directional sensor networks," in *Proc. IEEE INFOCOM*, Apr. 2009, pp. 2791–2795.

[19] W. Li, C. Huang, C. Xiao, and S. Han, "A heading adjustment method in wireless directional sensor networks," *Comput. Netw.*, vol. 133, pp. 33–41, Mar. 2018.

[20] S. Sharmin, F. N. Nur, M. A. Razzaque, and M. M. Rahman, "Area coverage for clustered directional sensor networks using Voronoi diagram," in *Proc. IEEE Int. WIE Conf. Electr. Comput. Eng. (WIECON-ECE)*, Dec. 2015, pp. 370–373.

[21] J. Chen, L. Zhang, and Y. Kuo, "Coverage-enhancing algorithm based on overlap-sense ratio in wireless multimedia sensor networks," *IEEE Sensors J.*, vol. 13, no. 6, pp. 2077–2083, Jun. 2013.

[22] S. Mirjalili, A. H. Gandomi, S. Z. Mirjalili, S. Saremi, H. Faris, and S. M. Mirjalili, "Salp swarm algorithm: A bio-inspired optimizer for engineering design problems," *Adv. Eng. Softw.*, vol. 114, pp. 163–191, Dec. 2017.

[23] S. Mirjalili, "The ant lion optimizer," *Adv. Eng. Softw.*, vol. 83, pp. 80–98, May 2015.
[24] Q. Wen, X. Q. Zhao, Y. P. Cui, Y. P. Zeng, H. Chang, and Y. J. Fu, "Coverage enhancement algorithm for WSNs based on vampire bat and improved virtual force," *IEEE Sensors J.*, vol. 22, no. 8, pp. 8245–8256, Apr. 2022.
[25] Y. Yao, Y. Li, D. Xie, S. Hu, C. Wang, and Y. Li, "Coverage enhancement strategy for WSNs based on virtual force-directed ant lion optimization algorithm," *IEEE Sensors J.*, vol. 21, no. 17, pp. 19611–19622, Sep. 2021.
[26] X. Zhao, Y. Cui, C. Gao, Z. Guo, and Q. Gao, "Energy-efficient coverage enhancement strategy for 3-D wireless sensor networks based on a vampire bat optimizer," *IEEE Internet Things J.*, vol. 7, no. 1, pp. 325–338, Jan. 2020.
[27] S. Peng and Y. Y. Xiong, "An area coverage and energy consumption optimization approach based on improved adaptive particle swarm optimization for directional sensor networks," *Sensors*, vol. 19, no. 5, p. 1192, 2019.
[28] G. Cheng and H. Wei, "Virtual angle boundary-aware particle swarm optimization to maximize the coverage of directional sensor networks," *Sensors*, vol. 21, no. 8, p. 2868, Apr. 2021.
[29] Y. Yao, H. Liao, X. Li, F. Zhao, X. Yang, and S. Hu, "Coverage control algorithm for DSNs based on improved gravitational search," *IEEE Sensors J.*, vol. 22, no. 7, pp. 7340–7351, Apr. 2022.
[30] D. Tao, S. Tang, and L. Liu, "Constrained artificial fish-swarm based area coverage optimization algorithm for directional sensor networks," in *Proc. IEEE 10th Int. Conf. Mobile Ad-Hoc Sensor Syst.*, Oct. 2013, pp. 304–309.
[31] Y. D. Yao, Q. Wen, Y. P. Cui, and B. Z. Zhao, "Discrete army ant search optimizer-based target coverage enhancement in directional sensor networks," *IEEE Sensors Lett.*, vol. 6, no. 4, pp. 1–4, Apr. 2022.
[32] F. Fan et al., "Optimized coverage algorithm of wireless video sensor network based on quantum genetic algorithm," *J. Commun.*, vol. 36, no. 6, pp. 98–108, 2015.
[33] X. Zhu, J. Li, X. Chen, and M. Zhou, "Minimum cost deployment of heterogeneous directional sensor networks for differentiated target coverage," *IEEE Sensors J.*, vol. 17, no. 15, pp. 4938–4952, Aug. 2017.
[34] X. Zhao, F. Yang, Y. Han, and Y. Cui, "An opposition-based chaotic salp swarm algorithm for global optimization," *IEEE Access*, vol. 8, pp. 36485–36501, 2020.
[35] X. Cai, P. Wang, L. Du, Z. Cui, W. Zhang, and J. Chen, "Multi-objective three-dimensional DV-hop localization algorithm with NSGA-II," *IEEE Sensors J.*, vol. 19, no. 21, pp. 10003–10015, Jul. 2019.
[36] J. Ai and A. A. Abouzeid, "Coverage by directional sensors in randomly deployed wireless sensor networks," *J. Combinat. Optim.*, vol. 11, no. 1, pp. 21–41, Feb. 2006.
[37] J. A. Manubay and S. Powell, "Detection of prey odours underpins dietary specialization in a neotropical top-predator: How army ants find their ant prey," *J. Animal Ecol.*, vol. 89, no. 5, pp. 1165–1174, May 2020.

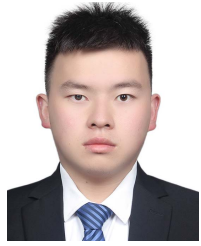

**Qin Wen** (Student Member, IEEE) received the M.S. degree from the Xi'an University of Posts and Telecommunications, Xi'an, China, in 2022. He is currently pursuing the Ph.D. degree with the School of Computer Science, Northwestern Polytechnical University (NPU), Xi'an.

His current research interests include the Internet of Things and wireless avionics intra-communications.

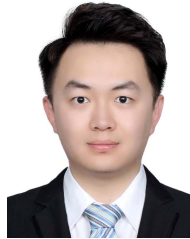

**Yan-Peng Cui** (Student Member, IEEE) received the B.S. degree from the Henan University of Technology, Zhengzhou, China, in 2016, and the M.S. degree from the Xi'an University of Posts and Telecommunications, Xi'an, China, in 2020. He is currently pursuing the Ph.D. degree with the School of Information and Communication Engineering, Beijing University of Posts and Telecommunications (BUPT), Beijing, China.

His current research interests include flying ad hoc networks, and integrated sensing and communication for unmanned aerial vehicle (UAV) networks.

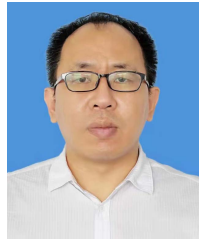

**Feng Zhao** is an Associate Professor with the Xi'an University of Posts and Telecommunications, Xi'an, China. His research interests include Internet of Things security and commercial password security evaluation.

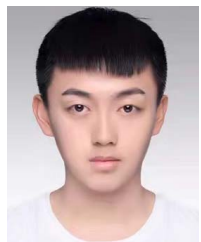

**Bo-Zhan Zhao** was born in 1999. He is currently pursuing the master's degree with the Xi'an University of Posts and Telecommunications, Xi'an, China.

His research interests include the technology and application of the Internet of Things.

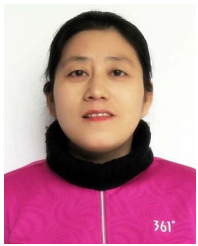

**Yin-Di Yao** was born in Baoji, China, in April 1978. She received the M.S. degree from the School of Computer Science, Shaanxi Normal University, Xi'an, China, in 2002.

She is a Senior Engineer with the Xi'an University of Posts and Telecommunications, Xi'an. Her current research interests include the technology and application of the Internet of Things.

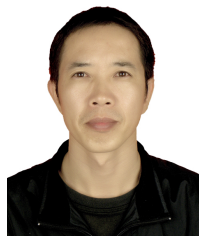

**Yao-Ping Zeng** was born in 1975.

He is an Associate Professor with the Xi'an University of Posts and Telecommunications, Xi'an, China. His research interests include wireless sensor optimization of network coverage and edge computing.